\documentclass[12pt]{article}
\title{Superconformal invariance from N=2 supersymmetry Ward identities}
\usepackage{mathrsfs}
\usepackage{amsmath}

\global\arraycolsep=1pt
\usepackage[colorlinks=true,backref=true,linkcolor=black,anchorcolor=black,citecolor=black,filecolor=black,menucolor=black,pagecolor=black,urlcolor=black]{hyperref}
\usepackage[Symbol]{upgreek}
\usepackage{bbm}
\usepackage{dsfont}
\usepackage{amssymb}
\usepackage{textcomp}
 
\newcommand{\baa}{/ \hspace{-1.4ex}}
\newcommand{\baaa}{\, / \hspace{-1.6ex}}

\newcommand{\Scal}[1]{\Bigl ({#1} \Bigr )}
\newcommand{\scal}[1]{\bigl ({#1} \bigr )}
\def\bea{\begin{eqnarray}}
\def\eea{\end{eqnarray}}
\def\be{\begin{equation}}
\def\ee{\end{equation}}
\newcommand{\CR}{\nonumber \\*}

\newcommand{\trace}{\hbox {Tr}~}

\def\L{{\ L}}

\def\spro{\, \overset{\centerdot}{\ }}
\def\repre{{\scriptscriptstyle \xi}}
\def\reprd{{\scriptscriptstyle \zeta}}
\def\ipro{\cdot}

\DeclareMathAlphabet{\mathpzc}{OT1}{pzc}{m}{it}
\def\s{\,\mathpzc{s}\,}
\def\a{{\scriptscriptstyle (\mathpzc{s})}}
\def\q{{{\scriptscriptstyle (Q)}}}
\def\qs{{\scriptscriptstyle (Q\mathpzc{s})}}

\def\L{{\cal L}}

\DeclareMathAlphabet{\mathpzc}{OT1}{pzc}{m}{it}
\def\s{\,\mathpzc{s}\,}

\def\a{{\scriptscriptstyle (\mathpzc{s})}}
\def\q{{{\scriptscriptstyle (Q)}}}
\def\qs{{\scriptscriptstyle (Q\mathpzc{s})}}

\def\Q{{\mathcal{S}_{\q}}}

\def\phis{{\phi^{\a}}}
\def\phiq{{\phi^{\q}}}
\def\phiqs{{\phi^{{\qs}}}}

\def\psiq{{\psi^{\q}}}
\def\psiqs{{\psi^{{\qs}}}}

\def\Omegas{{\Omega^{{\a}}}}
\def\Omegaq{{\Omega^{{\q}}}}
\def\Omegaqs{{\Omega^{{\qs}}}}

\def\cq{{c^{{\q}}}}
\def\muq{{\mu^{{\q}}}}

\def\A{{a}}

\def\S{{\mathcal{S}_\a}}

\def\epsilonb{{\overline{\epsilon}}}

\usepackage[Symbol]{upgreek}
\usepackage{caption}
\usepackage{bbm}
\usepackage{graphicx}

\def\ref{D\tilde}

\newcommand{\stfrac}[2]{{\textstyle \frac{#1}{#2}}}

\def\susy{{\delta^{\mathpzc{Susy}}}}
\def\conj{{\mathpzc{c}}}

\def\S{{\mathcal{S}_\a}}

\newcommand{\ins}[1]{\S_{|\Gamma} \bigl[
 \Uppsi^{\scriptscriptstyle{(#1)}} \ipro \Gamma \bigr]}

\def\CS{\mathcal{C}}

\def\L{{\cal L}}

\def\k{{\rm\scriptscriptstyle K}}
\def\bps{{\rm \scriptscriptstyle C}}
\def\hyp{{\rm \scriptscriptstyle H}}
\def\uk{{\, ^\k \hspace{-0.7mm} u}}
\def\uc{{\, ^\bps \hspace{-0.7mm} u}}
\def\vk{{\, ^\k \hspace{-0.7mm} v}}
\def\vc{{\, ^\bps \hspace{-0.7mm} v}}
\def\vh{{\, ^\hyp \hspace{-0.7mm} v}}
\def\ukb{{{\, ^\k \hspace{-0.7mm} \overline{u}}}}
\def\ucb{{{\, ^\bps \hspace{-0.7mm} \overline{u}}}}
\def\vkb{{{\, ^\k \hspace{-0.7mm} \overline{v}}}}
\def\vcb{{{\, ^\bps \hspace{-0.7mm} \overline{v}}}}
\def\vhb{{{\, ^\hyp \hspace{-0.7mm} \overline{v}}}}

\def\N{\mathcal{N}}

\begin{document}
\allowdisplaybreaks[1]
\renewcommand{\thefootnote}{\fnsymbol{footnote}}
\def\corr{$\spadesuit $}
\def\trefle{ $\clubsuit$}
\begin{titlepage}
\begin{flushright}
\
\vskip -3cm
{ \small AEI-2007-167}
\vskip 3cm
\end{flushright}
\begin{center}
{{\Large \bf
Superconformal invariance \\
\vskip 2mm
from $\N=2$ supersymmetry Ward identities}}
\lineskip .75em
\vskip 3em
\normalsize
{\large Laurent {\sc Baulieu}$^\dagger$ and Guillaume {\sc Bossard}\footnote{email address: bossard@aei.mpg.de}\\
\vskip 1 em
$^{\dagger}$ {\it Laboratoire Physique Th\'eorique et Hautes Energies\\ Universit\'e Pierre et Marie Curie\\ 4 Place Jussieu, 75005 Paris, France}
\\
\vskip 1 em
$^{*}$ {\it Max-Planck-Institut f\"{u}r Gravitationsphysik\\ Albert-Einstein-Institut\\ Am M\"{u}hlenberg 1, D-14476 Potsdam, Germany}
\\
}

\vskip 1 em
\end{center}
\vskip 1 em
\begin{abstract}
We algebraically prove the cancellation of the $\beta$ function at all order of perturbation theory of $\N=2$ supersymmetric gauge theories with a vanishing one-loop $\beta$ function. 
The proof generalises that recently given for the $\N=4$ case. It uses the consistent Slavnov--Taylor identities of the shadow dependent formulation. We also demonstrate the cancellation at all orders of the anomalous dimensions of vector and hypermultiplet \textonehalf BPS operators.
\end{abstract}

\end{titlepage}
\renewcommand{\thefootnote}{\arabic{footnote}}
\setcounter{footnote}{0}



\renewcommand{\thefootnote}{\arabic{footnote}}
\setcounter{footnote}{0}



\section{Introduction}
Many supersymmetric Yang--Mills theories are believed to admit as symmetry a larger set of superconformal generators. The superconformal symmetry strongly constrains the correlation functions of local composite operators. This permits the introduction of new methods of computations, beyond the usual rules of perturbation theory. For instance, in the maximally supersymmetric theory, the anomalous dimensions can be computed from an integrable Heisenberg spin chain model till three-loops \cite{Staudacher}. It has been conjectured that an integrable Heisenberg spin chain model exists which would permit to compute the anomalous dimensions of the model at all orders, and moreover, that this property can be extended to $\N=1$ superconformal deformations of the maximally supersymmetric theory. The superconformal Yang--Mills theories have been also conjectured to be duals to type IIB super-strings theories in Anti-de-Sitter like backgrounds \cite{AdS}.

The superconformal invariance is a consequence of supersymmetry and conformal invariance. One understands that the conformal invariance follows from the vanishing of the trace anomaly, a property that is closely linked to the cancellation of the $\beta$ function \cite{trace}. Using $\N=2$ harmonic superspace methods, it has been shown that the $\beta$ function can only receive one-loop contributions in perturbation theory \cite{harmonic}. It is in fact well admitted that all Yang--Mills theories with extended supersymmetry and a zero one-loop $\beta$ function admit a superconformal phase. 

Nevertheless, harmonic superspace path integrals are well-defined only if one use a regularization scheme that preserves gauge invariance and supersymmetry. However, such a regulator has not yet been defined, and the widely used dimensional reduction of Siegel can not preserve supersymmetry beyond three-loops \cite{consistence}.


It is of course interesting to demonstrate supercorformal invariance from first principles, using the locality property in quantum field theory and the symmetries of the lagrangian. Algebraic methods are well-suited for showing that there must exist a renormalization prescription for which the $\beta$ function is exact at one-loop in $\N=2$ super-Yang--Mills theory \cite{Sorella}. Here, we demonstrate
 that the all-order cancellation of the $\beta$ function of the $\N=2$ super-Yang--Mills theory is a consequence of its one-loop vanishing, independently of the regularization scheme. The proof uses two consistent Slavnov--Taylor identities respectively associated to gauge invariance and supersymmetry, which can be defined thanks to the introduction of shadow fields \cite{shadow}. It extends that proposed in \cite{beta} for the $\N=4$ super-Yang--Mills theories. Moreover, our earlier algebraic proof for the $\N=4$ case was formulated in function of twisted variables, which corresponds to a restriction of the manifest global symmetry invariance of the supersymmetric theory. Here, we will work directly in Minkowski space, making manifest the whole set of global symmetries. In this way, we avoid the subtile analytic deformations involved by the Wick rotation and the twist procedure.

The proof that the beta function vanishes amounts to show that the pure Yang--Mills lagrangian has no anomalous dimension and no possible mixing under renormalization with others operators than pure derivatives and BRST-exact terms. This property turns out to be a consequence of the cancellation at all orders of perturbation theory of the anomalous dimensions of the dimension $2$ and the dimension $\frac{5}{2}$ \textonehalf BPS operators of the vector multiplet. We prove the cancellation of the anomalous dimensions of all the \textonehalf BPS primary operators of the vector multiplet and of some \textonehalf BPS operators of the hypermultiplet. These results give further support for the superconformal invariance of these theories. The cancellation of the $\beta$ function is a main step in a proof of the superconformal invariance. Nevertheless, one has in principle to show that the whole set of physical composite operators belong to representations of the superconformal algebra. The cohomology elements of the linearized superconformal Slavnov--Taylor operator of shadow number two, linear in the external sources coupled to the physical operators, define possible obstructions for them to belong to superconformal representations at all order of perturbation theory.

\section{The action and its invariances}
We will consider an $\N=2$ super-Yang--Mills theory where the matter is such that the one-loop $\beta$ function is zero. 

The vector multiplet is in the adjoint representation of a simple compact gauge group $G$.\footnote{The generalization to a semi-simple compact gauge group is straightforward.} It is made of a gauge field $A_\mu$, one scalar $\phi$, one pseudo-scalar $\phi_5$, a fermionic $SU(2)$-Majorana spinor field $\lambda$, and an auxiliary field $H^i$ in the adjoint representation of the internal symmetry group $SU(2)_R$. These fields satisfy the following supersymmetry transformations\footnote{Our conventions are such that $\gamma_5^{\, 2} = -1$, $\tau^i$ are the Pauli matrices generating $SU(2)_R$ and $H\equiv H^i \tau_i$.} with a commuting $SU(2)$-Majorana parameter $\epsilon$
\begin{gather}\begin{split}
\susy A_\mu &= i\scal{\epsilonb \gamma_\mu \lambda} \\
\susy \phi &= - \scal{\epsilonb \lambda }\\
\susy \phi_5 &= - \scal{\epsilonb \gamma_5 \lambda }
\end{split}\hspace{10mm}\begin{split}
\susy \lambda &= \bigl( \baa F + i \baaa D (\phi + \gamma_5 \phi_5 )+ \gamma_5 [\phi, \phi_5] + H \bigr ) \epsilon \\
\susy H^i &= - i \scal{ \epsilonb \tau^i \baaa D \lambda} + \scal{ \epsilonb \tau^i [\phi + \gamma_5 \phi_5, \lambda]}
\end{split}\end{gather}
The matter fields belong to an hypermultiplet valued in a vector space on which the gauge group acts through a possibly reducible representation such that we can define an invariant scalar product.\footnote{All our statements extend to the case where the hypermultiplet scalar fields are valued in a quotient space of $\mathds{R}^{ (3+1)n}$ by any discrete subgroup of $SO(3)$.} 

 The Dynkin index of the hypermultiplet representation is chosen equal to the Dynkin index of the adjoint representation, so that the one-loop $\beta$ function be zero at first order. In the case of an $SU(N)$ gauge group, the hypermultiplet can be for example in the adjoint representation, with the trace as scalar product, or in the direct sum representation of $2N$ copies of the fundamental representation, with the antilinear isomorphism to the antifundamental representation as scalar product. 
 
As for the $SU(2)_R$ R-symmetry assignments, the hypermultiplet is made of one singlet scalar $L$, one triplet of scalars $h^i$, and of an $SU(2)$-Majorana spinor $\psi$. One has 
\begin{gather}\begin{split}
\susy L &= - \scal{\epsilonb \psi }\\
\susy h^i &= - i \scal{\epsilonb\tau^i \psi }
\end{split}\hspace{10mm}\begin{split}
\susy \psi = \bigl[ i \baaa D L + \baaa D h -\phi L - \gamma_5 \phi_5 L + i \phi h + i \gamma_5 \phi_5 h \bigr] \epsilon 
\end{split}\end{gather}
The pure Yang--Mills and matter supersymmetric Lagrange densities are respectively 
\begin{multline}
\L_{\rm YM} \equiv \trace \Bigl(- \stfrac{1}{4} F_{\mu\nu} F^{\mu\nu} -
\stfrac{1}{2} D_\mu \phi D^\mu \phi - \stfrac{1}{2} D_\mu \phi_5 D^\mu \phi_5 + \stfrac{i}{2}
\scal{\overline{\lambda} \baaa D \lambda } \\*- \stfrac{1}{2} \scal{
\overline{\lambda} [ \phi, \lambda]} - \stfrac{1}{2} \scal{
\overline{\lambda} \gamma_5 [ \phi_5, \lambda]}- \stfrac{1}{2}
[\phi,\phi_5]^2- \stfrac{1}{2} H^i H_i \Bigr) 
\end{multline}
\begin{multline}
\L_{\rm H} \equiv -
\stfrac{1}{2} D_\mu L \spro D^\mu L- \stfrac{1}{2} D_\mu h^i \spro D^\mu h_i + \stfrac{i}{2}
\scal{\overline{\psi}\spro \baaa D \psi } \\* + \stfrac{1}{2} \scal{
\overline{\psi} \spro \phi \psi} - \stfrac{1}{2} \scal{\overline{\psi} \spro \gamma_5 \phi_5 \psi} +\scal{
\overline{\psi} \spro \lambda} L - i \scal{\overline{\psi} \spro \tau_i \lambda} h^i \\*+ \stfrac{1}{2} L \spro \phi^2 \, L + \stfrac{1}{2} L \spro \phi_5^{\, 2} \, L + \stfrac{1}{2} h^i \spro \phi^2 \, h_i + \stfrac{1}{2} h^i \spro \phi_5^{\, 2} \, h_i + i L \spro H^i h_i + \stfrac{i}{2} \varepsilon_{ijk} h^i \spro H^j h^k \label{hyper}
\end{multline}
Here the symbol ``$\spro $" denotes the $G$-invariant scalar product between the fields of the hypermultiplet.

To achieve a supersymmetric and BRST invariant gauge-fixing, we introduce shadow fields $c$ and $\mu$ and the Faddeev--Popov ghost field $\Omega$ \cite{shadow}.
These scalar fields belong to the adjoint representation of the gauge group. The shadow and ghost numbers of $c$, $\mu$ and $\Omega$ are $(1,0)$, $(1,1)$ and $(0,1)$, respectively.

The action of the BRST operator $\s$ on the physical fields, (i.e, all fields contained in the vector multiplet and the hypermultiplet) is identical to a gauge transformation with parameter $-\Omega$. The action of the differential $Q$ on them is a supersymmetry transformation with parameter $\epsilon$ minus a gauge transformation of parameter $c$, $Q=\susy(\epsilon) - \delta^{\rm gauge }(c) $. For the unphysical fields $\Omega$, $c$ and $\mu$, the action of $\s$ and $Q$ is
\begin{gather}\begin{split}
\s \Omega &= - \Omega^2 \\
\s c &= \mu\\
\s \mu &= 0
\end{split}\hspace{10mm}\begin{split}
Q \Omega &= - \mu - [c, \Omega] \\
Q c &= \scal{\epsilonb [ \phi + \gamma_5 \phi_5 - i \baaa A] \epsilon} - c^2 \\
Q \mu &= - (\epsilonb\epsilon) [\phi, \Omega] - (\epsilonb\gamma_5 \epsilon) [\phi_5, \Omega]+ i (\epsilonb \gamma^\mu \epsilon) D_\mu \Omega - [c, \mu]
\end{split}\end{gather}
The difficulties caused by the appearance of gauge transformations in the closure relations of the supersymmetry algebra are bypassed since one has 
\be \s^2= 0 \hspace{10mm} \{ \s, Q \} = 0 \hspace{10mm} Q^2 \approx - i (\epsilonb \gamma^\mu \epsilon) \partial_\mu \ee 
The symbol $ \approx $ means that the last equality holds modulo the equation of motion of the fermionic matter field $\psi$. The Batalin--Vilkovisky method permits one to go around this annoying property, by introducing a quadratic term in the sources that depends on $\epsilon$ in the complete action, which will be shortly made explicit in Eq.(8).

Antighosts and antishadows are needed for writing a $Q$-invariant and $\s$-exact gauge-fixing action with ghost and shadow number zero. We thus have the trivial quartet of fields $\bar\mu, \, \bar c ,\, \bar \Omega$ and $b$, with $Q$ and $\s$ transformations given in \cite{shadow}. The $\s$-exact and $Q$ invariant gauge-fixing action is
\be\s \Uppsi \equiv -\s Q \int d^4 x \trace \Scal{ \bar \mu
\, \partial^\mu A_\mu + \stfrac{\alpha}{2} \bar \mu b} \label{gaugefixing}\ee
One must also introduce all relevant sources for the non-linear $Q$ and $\s$ transformations of all fields. One ends up with the following complete action 
 \begin {multline} 
\Sigma \equiv \frac{1}{g^2} \int d^4 x \Scal{ \L_{\rm YM} + \L_{\rm H}} - \int d^4 x \trace \Bigl( b
\partial^\mu A_\mu - \bar \Omega \partial^\mu
D_\mu \Omega + \frac{\alpha}{2} b^2 \\* + \frac{i \alpha}{2} (\epsilonb \gamma^\mu \epsilon) \bar c
\partial_\mu \bar c + \bar c
\partial^\mu \scal{D_\mu c + i(\epsilonb \gamma_\mu \lambda)}
+ \bar \mu \partial^\mu \scal{D_\mu \mu + [D_\mu \Omega, c] - i
(\epsilonb \gamma_\mu [\Omega, \lambda])} \Bigr) \\*+ \frac{g^2}{4} \int d^4 x \overline{[\psiq - \psiqs \Omega]} \scal{ (\epsilonb \epsilon) + (\epsilonb \gamma_5 \epsilon) \gamma_5 - (\epsilonb \gamma^\mu \epsilon) \gamma_\mu } [ \psiq - \Omega \psiqs] \\* 
+ \int d^4 x (-1)^\A \Scal{ \phis_\A \s \varphi^\A +\phiq_\A Q \varphi^\A +
\phiqs_\A \s Q \varphi^\A} \\* 
+ \int d^4 x \trace \biggl( \Omegas \Omega^2 -
\Omegaq Q \Omega - \Omegaqs \s Q \Omega - \cq Q c + \muq Q \mu
 \biggr) \label{action}
\end{multline}
$\Sigma$ is invariant under all the global symmetries of the theory. It also satisfies both Slavnov--Taylor identities $\S(\Sigma) = 0$ and $\Q(\Sigma) =0$, respectively associated to BRST invariance and the $Q$ supersymmetry \cite{shadow}. In the class of linear gauges that we have chosen, one has also a quartet of antighost Ward identities, related to the equations of motion of the antighost fields. Moreover, when the gauge parameter $\alpha=0$, there are three additional ghost Ward identities \cite {shadow}.

Using algebraic methods, one can show that there is no consistent anomaly to the Ward identities and the Slavnov--Taylor identities in the case of a semi-simple gauge group \cite{shadow,White}. Thus, from the principle of locality, one concludes that there exists renormalization prescriptions that preserve all these Ward identities at the quantum level.
 
The non-trivial part of the $U(1)_A$ and the $SU(2)_R$ symmetries of the theory are defined by 
\be\begin{split}
\delta^A \lambda &= \omega \gamma_5 \lambda \\
\delta^R \lambda &= i \upsilon_i \tau^i \lambda 
\end{split}\hspace{10mm}\begin{split}
\delta^A \phi &= 2 \omega \phi_5 \\
\delta^R H^i &= 2 \upsilon_k {\varepsilon^{ki}}_j H^j
\end{split}\hspace{10mm}\begin{split}
\delta^A \phi_5 &= - 2 \omega \phi \\
\delta^R h^i &= 2 \upsilon_k {\varepsilon^{ki}}_j h^j 
\end{split}\hspace{10mm}\begin{split}
\delta^A \psi &= - \omega \gamma_5 \psi \\
\delta^R \psi &= i \upsilon_i \tau^i \psi 
\end{split}\ee
To determine the possible anomalies for the internal symmetry group $SU(2)_R\times U(1)_A$, we consider the supersymmetry Slavnov--Taylor identity for the enlarged 
 differential $Q \rightarrow Q+ \delta^A + \delta^R$,\footnote{with $\omega$ and $\upsilon_i$ defined as anticommuting parameters.} which results into the following parameter-dependent modification of the Slavnov--Taylor operator
\be \Q(\Sigma) + [ \omega \gamma_5 + i \upsilon_i \tau^i ] \epsilon \frac{\partial\, \Sigma}{\partial \epsilon} - {\varepsilon_i}^{jk} \upsilon_j \upsilon_k \frac{\partial\, \Sigma}{\partial \upsilon_i}=0 \ee
One then finds that the only consistent anomaly is determined by the cohomological element $\frac{\omega}{4\pi^2} \int \trace F_{\, \wedge} F$. At the perturbative level, if one remains around a trivial instantonic vacuum, this term is zero and the whole internal symmetry group is preserved by the radiative corrections. Although the axial symmetry $U(1)_A$ is preserved, the Ward identity associated to the conservation low of its courant is however usually broken at one-loop by a term proportional to $\frac{1}{4\pi^2} \int \omega(x) \trace F_{\, \wedge} F$. The coefficient of the anomaly is known to be exact at one-loop, and is proportional to the one-loop $\beta$ function. The axial courant Ward identity is thus preserved in the theories we are considering. 
\section{Finiteness of vector multiplet \textonehalf BPS primaries}
In \cite{beta} and \cite{BPS}, the question of the cancellation of the anomalous dimensions of the \textonehalf BPS primary operators of the vector multiplet was addressed to, by using the Slavnov--Taylor identities of euclidean supersymmetric theories in a twisted form (with all spinorial indices mapped on tensor indices).

Here, we wish to work in four-dimensional Minkowski spacetime, with all fermionic physical fields expressed as spinors. As we will see shortly, the Minkowski formulation exhibits the fact that an analytic continuation of the supersymmetry parameter that defines the Slavnov--Taylor identities is required by the proof. Indeed, this permits a reduction of the number of parameters of supersymmetry transformations with a Lorentz and $R$ symmetry fully invariant constraint. Let us be more precise on this.

Given an $SU(2)$-Majorana spinor $\epsilon$, its conjugated $\epsilonb$ is equal to the spinor itself, contracted with the $SU(2)$ charge conjugation matrix, written $\epsilon^\conj$. In the supersymmetric gauges of \cite{shadow}, the propagators and vertex are polynomials in the supersymmetry parameter $\epsilon$, and because of the shadow number conservation, the generating functional of one particle irreducible graphs $\Gamma$ is a polynomial in $\epsilon$. 
 
 This regularity permits one to analytically continue the generating functional $\Gamma$ to arbitrary complex values of $\epsilon$ in such way that all the Ward identities, including the Slavnov--Taylor identities, still hold. The supersymmetry Slavnov--Taylor identity corresponds no longer to an invariance of the theory by ``infinitesimal" supersymmetry shift of the fields, but to formal Ward identities, like the supersymmetry Ward identities in the euclidean formulation of the theory, where the parameters have become complex.
 
 The gain of this analytic continuation is as follows. Given an $SU(2)$-Majorana spinor $\epsilon$, the only solution of the equation $(\epsilonb \gamma^\mu \epsilon) = 0$ is $\epsilon=0$. On the other hand, when $\epsilon$ is complex, one has non-trivial solution for the equation 
\be \epsilon^\conj \gamma^\mu \epsilon = 0 \label{purespinor}\ee
Then, when the last condition is satisfied, the Fierz identity implies the following equation
\be (\epsilon^\conj \epsilon)^2 + (\epsilon^\conj \gamma_5 \epsilon)^2= 0\ee
In turn, it implies that $\Phi \equiv (\epsilon^\conj \epsilon) \phi + (\epsilon^\conj \gamma_5 \epsilon) \phi_5 $ is proportional to the first component of a chiral (or antichiral) superfield $\phi \pm i \phi_5$, for an $\N=1$ supersymmetry subalgebra of the whole $\N=2$ one. Choosing a solution of (\ref{purespinor}) that verifies either $(\epsilon^\conj \epsilon)= i (\epsilon^\conj \gamma_5\epsilon)$ or
$(\epsilon^\conj \epsilon)=- i (\epsilon^\conj \gamma_5 \epsilon)$, one gets that the invariant polynomials of $\Phi$ are either propotional to the invariant polynomials
 of $\phi - i \phi_5$ or of $\phi + i \phi_5$, respectively. In this way, one obtains the whole set of \textonehalf BPS primary operators of the vector multiplet.

We can compute all possible insertions of invariant polynomials $\mathcal{P}(\Phi)$ 
in the Green functions of the theory, expressed in the supersymmetric gauge defined by the action (\ref{gaugefixing}), with a value of the gauge parameter $\alpha$ set equal to zero and a supersymmetric parameter $\epsilon$ satisfying (\ref{purespinor}). According to the above reasoning, these insertions depend in a polynomial way on the complex supersymmetry parameter $\epsilon$. The result is fully compatible with the global $Spin(3,1)\times U(1)_A \times SU(2)_R$ symmetry. The general factorization property for physical composite operators \cite{beta} warrantees that these insertions are equivalent, up to irrelevant BRST-exact insertions, to insertions of the \textonehalf BPS primary operators.

Then, the point is that, due to the constraint $ \epsilon^\conj \gamma^\mu \epsilon = 0$, one has 
$Q c + c^2= \Phi $, so that, by using the Chern--Simons formula, any given 
 invariant polynomial in $\Phi$ can be written as the $Q$ variation of a composite operator $ \Delta(c,\Phi)$, which involve the shadow field $c$.

 By a mere adaptation of the proof \cite{beta,BPS}, one can then show that the ghost Ward identities valid in the Landau gauge $(\alpha =0)$ imply that the composite operators $ \Delta(c,\Phi)$ have zero anomalous dimensions, and then the supersymmetry Slavnov--Taylor identity imply that the composite operators $\mathcal{P}(\Phi)$ have zero anomalous dimensions too.

Moreover, since the \textonehalf BPS primary operators are non-trivial cohomological elements of the linearized BRST Slavnov--Taylor operator as invariant polynomials, and since their $U(1)_A$ representations and canonical dimensions protect them from being possibly mixed by renormalization with other operators, they are finite operators for any value of the gauge parameters $\alpha$ and $\epsilon$. This means that they are left invariant by the the Callan--Symanzik functional operator $\ \CS $,
\be \CS \Bigl[\, \mathcal{P}(\phi \pm i \phi_5 ) \ipro \Gamma \, \Bigr] = 0 \ee
\section{Solutions of the descent equations}

Both terms $\L_{YM} $ and $ \L_{H } $ of the lagrangian are Hodge dual to gauge invariant four-forms, function of the fields and invariant with respect with supersymmetry up to an exterior derivative term, which we can denote generically as $ \L_4^0$.

The condition $\susy \L_4^0 =- d \L_3^1$ implies from the algebraic Poincar\'e lemma that the 3-form $\L_3^1$ verifies $\susy \L_3^1 =- d \L_2^2 - i_{i (\epsilonb \gamma \epsilon)} \L_4^0$ and so on till zero form degree \cite{descent}. 
It is convenient to define the formal sum of differential forms of form degree plus shadow number equal to four, 
$ \tilde \L_4= \L_4^0 + \L_3^1+ \L_2^2 + \L_1^3 + \L_0^4 $.

 We will compute the non-trivial solutions of the descent equations of canonical dimension four, defined as non-trivial cohomological elements with zero ghost number of the linearized BRST Slavnov--Taylor operator $\S_{|\Sigma}$ and non-trivial cohomological elements of the extended nilpotent differential $d + \Q_{|\Sigma} + i_{i (\epsilonb \gamma \epsilon)}$. These elements are in one to one correspondence with the gauge invariant $4$-forms functions of the physical fields, invariant under the supersymmetry generators up to an exterior derivative term.

 One easily obtains a cocycle corresponding to the second Chern character thanks to the curvature equation that reproduces the supersymmetry transformations\footnote{with the notations $\gamma_{\mathpzc{1}} \equiv \gamma_\mu dx^\mu$ and $\gamma_{\mathpzc{2}} \equiv \stfrac{1}{2} \gamma_{\mu\nu} dx^\mu_{\, \wedge}dx^\nu$, that is $- i \scal{\epsilonb \gamma_{\mathpzc{1}} \lambda} \equiv i \scal{\epsilonb \gamma_\mu \lambda} dx^\mu$.}
\be (d + Q + i_{i (\epsilonb \gamma \epsilon)}) \scal{A + c} + \scal{A + c}^2 = F - i \scal{\epsilonb \gamma_{\mathpzc{1}} \lambda} + \scal{\epsilonb [\phi + \gamma_5 \phi_5 ] \epsilon}\ee
Indeed, this equation indicates that the following expression is left invariant by $ (d + Q + i_{i (\epsilonb \gamma \epsilon)})$, and is therefore a cocycle
\be
 \trace \Scal{ F - i \scal{\epsilonb \gamma_{\mathpzc{1}} \lambda} + \scal{\epsilonb [\phi + \gamma_5 \phi_5 ] \epsilon}}_{\, \wedge }\Scal{ F - i \scal{\epsilonb \gamma_{\mathpzc{1}} \lambda} + \scal{\epsilonb [\phi + \gamma_5 \phi_5 ] \epsilon}} \ee
We will however prefer the equivalent cohomological class 
\begin{multline}
 \tilde \L_{{\rm Ch}}\equiv \trace \Scal{ F - i \scal{\epsilonb \gamma_{\mathpzc{1}} \lambda} + \scal{\epsilonb [\phi + \gamma_5 \phi_5 ] \epsilon}}_{\, \wedge }\Scal{ F - i \scal{\epsilonb \gamma_{\mathpzc{1}} \lambda} + \scal{\epsilonb [\phi + \gamma_5 \phi_5 ] \epsilon}} \\*+ (d + Q + i_{i (\epsilonb \gamma \epsilon)}) \trace \Scal{ 2 \scal{\epsilonb \gamma_{\mathpzc{2}} \phi \lambda} + 2 \scal{\epsilonb\gamma_5 \gamma_{\mathpzc{2}} \phi_5 \lambda } + \stfrac{i}{2} (\epsilonb \gamma_{\mathpzc{1}} \epsilon) \scal{ \phi^2 + \phi_5^{\, 2}}} \end{multline}
for which the two last components of form degree $1$ and $0$ are \textonehalf BPS operators.

The others cocycles must be computed explicitly. This computation is made easier by solving the Hodge dual version of the descent equations \cite{descent}
\bea
\Q_{|\Sigma} \L + \partial^\mu \L_\mu &=& \ \ 0 \CR
\Q_{|\Sigma} \L_\mu + \partial^\nu \L_{\mu\nu} &=& \ \ i (\epsilonb \gamma_\mu \epsilon ) \L \CR
\Q_{|\Sigma} \L_{\mu\nu} + \partial^\sigma \L_{\mu\nu\sigma} &=&- 2 i (\epsilonb \gamma_{[\mu} \epsilon ) \L_{\nu]} \CR
 \Q_{|\Sigma} \L_{\mu\nu\sigma} + \partial^\rho \L_{\mu\nu\sigma\rho} &=&\ \ 3 i (\epsilonb \gamma_{[\mu} \epsilon ) \L_{\nu\sigma]} \CR
 \Q_{|\Sigma} \L_{\mu\nu\sigma\rho} &=& - 4 i (\epsilonb \gamma_{[\mu} \epsilon ) \L_{\nu\sigma\rho]} 
 \eea
For the components associated to $\L_{\rm YM } $, one gets 
 \begin{multline}
 \L_{\rm YM\ \mu } =
\trace\biggl(\frac{i}{2} F_{\mu\nu} \scal{ \epsilonb \gamma^\nu \lambda } + \frac{i}{4} {\varepsilon_{\mu\nu}}^{\sigma\rho} F_{\sigma\rho} \scal{ \epsilonb \gamma^\nu \lambda } - \frac{1}{2} D_\mu \phi \scal{ \epsilonb \lambda } - \frac{1}{2} D_\mu \phi_5 \scal{ \epsilonb \gamma_5 \lambda } \\* + \phi \scal{ \epsilonb {\gamma_\mu}^\nu D_\nu \lambda } + \phi_5 \scal{ \epsilonb\gamma_5 {\gamma_\mu}^\nu D_\nu \lambda } - \frac{i}{2} [\phi, \phi_5] \scal{ \epsilonb\gamma_5 \gamma_\mu \lambda } - \frac{i}{2} \Scal{ \epsilonb \gamma_\mu H \lambda} \biggr)
\end{multline}
\begin{multline}
 \L_{\rm YM\ \mu\nu } =
 \trace \biggl(\frac{1}{4} {\varepsilon_{\mu\nu\sigma}}^{\rho} (\epsilonb \gamma^\sigma \epsilon ) \scal{ \overline{\lambda} \gamma_5 \gamma_\rho \lambda } + \frac{1}{2} (\epsilonb \gamma_5 \epsilon )\Scal{ \phi_5 F_{\mu\nu} +\stfrac{1}{2} {\varepsilon_{\mu\nu}}^{\sigma\rho} \phi F_{\sigma\rho} + \scal{ \overline{\lambda} \gamma_5 \gamma_{\sigma\rho} \lambda }} \\*+ \frac{1}{2} (\epsilonb \epsilon )\Scal{ \phi F_{\mu\nu} - \stfrac{1}{2} {\varepsilon_{\mu\nu}}^{\sigma\rho} \phi_5 F_{\sigma\rho} + \scal{ \overline{\lambda} \gamma_{\sigma\rho} \lambda }} -(\epsilonb \gamma_{\mu\nu} \tau_i \epsilon ) \Scal{ \phi H^i - \frac{1}{2} \scal{ \overline{\lambda} \tau^i \lambda }} \\*- \frac{1}{2} {\varepsilon_{\mu\nu}}^{\sigma\rho} (\epsilonb \gamma_{\sigma\rho} \tau_i \epsilon ) \Scal{ \phi_5 H^i - \frac{1}{2} \scal{ \overline{\lambda} \gamma_5 \tau^i \lambda }}
\biggr) - \frac{i}{2} (\epsilonb \gamma_{[\mu} \epsilon) \partial_{\nu]} \trace \scal{ \phi^2+ \phi_5^{\, 2}}
\end{multline}
\be \L_{\rm YM\ \mu\nu\rho } =
- \frac{i}{2} \varepsilon_{\mu\nu\sigma\rho} \bigr[ ( \epsilonb \epsilon ) \epsilonb \gamma_5 \gamma^\rho + ( \epsilonb \gamma_5 \epsilon ) \epsilonb \gamma^\rho \bigr] \, \trace \scal{ \, [\phi-\gamma_5 \phi_5] \lambda }
\ee
\be
 \L_{\rm YM\ \mu\nu\rho \sigma} =
 \frac{1}{2} \varepsilon_{\mu\nu\sigma\rho} \Scal{ ( \epsilonb \epsilon ) ( \epsilonb \gamma_5\epsilon ) \trace \scal{\phi_5^{\, 2} - \phi^2} + \scal{ (\epsilonb \epsilon)^2 - (\epsilonb \gamma_5 \epsilon)^2 } \trace \scal{\phi \phi_5 }}\ee
 
For the hypermultiplet density, there is a difficulty because the supersymmetry transformations of its fermions close only modulo equations of motion. However, in a way analogous as in \cite{Losev}, we can introduce the $\epsilon$-dependent field $G(\epsilon) \equiv \frac{g^2}{2} [ (\epsilonb\gamma^\mu \epsilon) \gamma_\mu - (\epsilonb \epsilon)- (\epsilonb \gamma_5 \epsilon) \gamma_5 ] (\psiq - \Omega \psiqs)$, which plays the role of an auxiliary field, as can be verified from the following equations 
\bea
 \Q_{|\Sigma} \psi &=& \bigl[ i \baaa D L + \baaa D h -\phi L - \gamma_5 \phi_5 L + i \phi h + i \gamma_5 \phi_5 h \bigr] \epsilon + G(\epsilon)- c \, \psi \\*
 \Q_{|\Sigma} G(\epsilon)&=& \frac{1}{2} [ (\epsilonb\gamma^\mu \epsilon) \gamma_\mu - (\epsilonb \epsilon)- (\epsilonb \gamma_5 \epsilon) \gamma_5 ] [ - i \baaa D \psi - \phi \psi + \gamma_5 \phi_5 \psi - \lambda L + i \tau_i \lambda h^i ] - c \, G(\epsilon) \nonumber \eea
As compared to the interpretation given in \cite{Losev}, we simply consider $G(\epsilon)$ as a parameter dependent combination of sources and Faddeev--Popov ghosts, which permits one 
 to impose the analyticity in the supersymmetry parameter $\epsilon$. 
 
It must be noted that $G(\epsilon)$ satisfies the following constraints
\be{ \epsilonb\, G(\epsilon)}=0\hspace{10mm} { \epsilonb \tau^i \, G(\epsilon)} = 0\ee
Thus, it only counts for the four degrees of freedom that are required for the balance between the fermionic and the bosonic degrees of freedom of the hypermultiplet.

In order to make the hypermultiplet Lagrange density (\ref{hyper}) invariant under the action of $\Q_{|\Sigma}$ up to a derivative term, one must add to $\L_{H}$ the following BRST-exact term
\be - \frac{1}{2 (\epsilonb\gamma^\nu \epsilon)(\epsilonb\gamma_\nu \epsilon)} (\epsilonb\gamma_\mu \epsilon) \scal{\overline{G(\epsilon)} \spro \gamma^\mu G(\epsilon)} = \S_{|\Sigma} \, \frac{g^2}{2} \scal{ \overline{G(\epsilon)} \spro \psiqs} \ee 
This term is nothing but the rewriting in function of $G(\epsilon)$ of the quadratic term in the sources of the action (\ref{action}), which is analytic in the supersymmetry parameter $\epsilon$.

Given any $G$-invariant symmetrical rang two tensor $T^\repre$ of the hypermultiplet representation, we can now compute associated cocycles. Using the notation $\spro^\repre$ for the contraction of the indices in the representation of the gauge group with the invariant tensor $T^\repre$, the components of the cocycles are
\begin{multline}
 \L^\repre_{\rm H\ \mu } =
- \frac{1}{2} D_\mu L \spro^\repre \, \scal{\epsilonb \psi} - \frac{i}{2} D_\mu h^i \spro^\repre \, \scal{\epsilonb \tau_i \psi} - D^\nu L \spro^\repre \, \scal{\epsilonb \gamma_{\mu\nu} \psi } - i D^\nu h^i \spro^\repre \, \scal{\epsilonb \gamma_{\mu\nu} \tau_i \psi } \\* - \frac{i}{2} L \spro^\repre \, \phi \scal{\epsilonb \gamma_\mu \psi} - \frac{i}{2} L \spro^\repre \, \phi_5 \scal{\epsilonb \gamma_5 \gamma_\mu \psi} + \frac{1}{2} h^i \spro^\repre \, \phi \scal{\epsilonb \gamma_\mu \tau_i \psi} + \frac{1}{2} h^i \spro^\repre \, \phi_5 \scal{\epsilonb \gamma_5 \gamma_\mu \tau_i \psi} \\*- L \spro^\repre \, \scal{\epsilonb \gamma_\mu \tau_i \lambda } h^i - \frac{1}{2} \varepsilon_{ijk} h^i \spro^\repre \, \scal{ \epsilonb \gamma_\mu \tau^j \lambda} h^k + \frac{i}{2} \scal{\overline{\psi} \spro^\repre \, \gamma_\mu G(\epsilon)} + 2 \partial^\nu \scal{\epsilonb \gamma_{\mu\nu} [ L + \stfrac{i}{3} h ]\spro^\repre \, \psi }
\end{multline}
\begin{multline}
 \L^\repre_{\rm H\ \mu\nu } =
{\varepsilon_{\mu\nu\sigma}}^\rho (\epsilonb \gamma_5 \gamma^\sigma \tau_i \epsilon) \scal{ \stfrac{1}{2} {\varepsilon^i}_{jk} h^j \spro^\repre \, D_\rho h^k - L\spro^\repre \, D_\rho h^i } + 2i (\epsilonb \gamma_{\mu\nu} \tau_i \epsilon) L \spro^\repre \, \phi h^i \\*+ 2i (\epsilonb \gamma_5 \gamma_{\mu\nu} \tau_i \epsilon) L \spro^\repre \, \phi_5 h^i - i \varepsilon_{ijk} (\epsilonb \gamma_{\mu\nu} \tau^i \epsilon ) h^j \spro^\repre \, \phi h^k - i \varepsilon_{ijk} (\epsilonb \gamma_5 \gamma_{\mu\nu} \tau^i \epsilon ) h^j \spro^\repre \, \phi_5 h^k \\*- \frac{1}{4} \varepsilon_{\mu\nu\sigma\rho} (\epsilonb \gamma^\sigma \epsilon) \scal{ \overline{\psi} \spro^\repre \, \gamma_5 \gamma^\rho \psi} - \frac{1}{2} (\epsilonb \epsilon) \scal{ \overline{\psi} \spro^\repre \, \gamma_{\mu\nu} \psi} + \frac{1}{2} ( \epsilonb \gamma_5 \epsilon) \scal{ \overline{\psi} \spro^\repre \, \gamma_5 \gamma_{\mu\nu} \psi} \\*
- 2 \susy \scal{\epsilonb \gamma_{\mu\nu} [ L + \stfrac{i}{3} h] \spro^\repre \, \psi }- 2 \scal{ \epsilonb \gamma_{\mu\nu} [ L + \stfrac{i}{3} h] \spro^\repre \, G(\epsilon) }
\end{multline}
\be \L^\repre_{\rm H\ \mu\nu\rho } = \frac{i}{2} \varepsilon_{\mu\nu\sigma\rho} \, \varepsilon_{ijk} ( \epsilonb \gamma_5 \gamma^\rho \tau^i) \scal{ \epsilonb \tau^j [ h^k - \stfrac{1}{3} \tau^k h ] \spro^\repre \, \psi } \ee
and the last form $ \L^\repre_{\rm H\ \mu\nu\rho\sigma } $ is zero.

Notice that we have adjusted the cocycles associated respectively to the vector multiplet and the hypermultiplet in such way that the corresponding forms of form degree one and shadow number three are composite operators belonging to \textonehalf BPS multiplets. These are the simplest expressions that we can reach by adjusting the representatives of the corresponding cohomology class. 
\section{Anomalous dimensions and Ward identities}
To compute and adjust the correlation functions involving composite operators, one usually include terms in the action with external sources coupled to them. Then, the action becomes $\Sigma[u] = \Sigma +\Upsilon[u]$. In order to preserve the supersymmetry Slavnov--Taylor identity, one has to introduce complete supermultiplets of composite operators and to define the transformations of the sources. Let us introduce the simplest operators
\begin{multline}
\Upsilon[u] \equiv \int d^4 x \biggl( \uc\, \stfrac{1}{2}\trace \scal{ \phi_5^{\, 2} - \phi^2 } + \uc_5 \trace \phi \phi_5 + \ucb \trace [ \phi - \gamma_5 \phi ] \lambda \\* + \uk \, \stfrac{1}{2} \trace \scal{ \phi^2 + \phi_5^{\, 2} } + \ukb \trace [ \phi + \gamma_5 \phi_5 ] \lambda + \vc^\repre_{ij} \, \stfrac{1}{2} \scal{ h^i \spro_\repre h^j - \stfrac{1}{3} \delta^{ij} h^k \spro_\repre h_k } \\*+ \vcb_i^\repre [h^i - \stfrac{1}{3} \tau^i h ] \spro_\repre \psi + \vk^\repre \, \stfrac{1}{2} \scal{ L \spro_\repre L + \stfrac{1}{3} h^i \spro_\repre h_i } + \vkb^\repre [ L + \stfrac{i}{3} h ] \spro_\repre \psi \\* + \vh^\repre_i L \spro_\repre h^i + \vh^\repre \, \stfrac{1}{2} \scal{ L \spro_\repre L - \stfrac{1}{3} h^i \spro_\repre h_i } + \vhb^\repre [ L - \stfrac{i}{3} h ] \spro_\repre \psi + \cdots \biggr)
\end{multline}
Here the dots stand for higher dimensional operators of the same multiplets that include non-physical BRST-exact composite operators depending of $G(\epsilon)$, since the supersymmetry representation is only defined modulo the equations of motion. The supersymmetry Slavnov--Taylor identity is now extended as follows
\begin{multline}
\Q(\Sigma) + \Q_{|\Sigma} \Upsilon[u] + \int d^4 x\biggl( Q u^I \frac{\partial^L \Upsilon[u]}{\partial u^I} \biggr) \\*+ \frac{g^2}{2} \int d^4x \biggl( \frac{\delta^R \Upsilon[u]}{\partial \psi} \spro \bigl[ (\epsilonb \gamma^\mu \epsilon ) \gamma_\mu - (\epsilonb \epsilon) - (\epsilonb \gamma_5 \epsilon) \gamma_5 \bigr] \frac{\delta^L \Upsilon[u]}{\delta \overline{G(\epsilon)}} \biggr) =0
 \end{multline}
Neglecting the terms coming from sources coupled to higher dimensional operators, we have the transformations 
\begin{gather}
Q \uc = i [\gamma^\mu \epsilon]_\alpha \partial_\mu \ucb^\alpha \hspace{7mm} Q \uc_5 = i [\gamma_5 \gamma^\mu \epsilon]_\alpha \partial_\mu \ucb^\alpha \hspace{7mm} Q \ucb^\alpha = - [\epsilonb]^\alpha \uc + [\epsilonb \gamma_5 ]^\alpha \uc_5 + \cdots \CR
Q \uk = - i [\gamma^\mu \epsilon]_\alpha \partial_\mu \ukb^\alpha \hspace{20mm} Q \ukb^\alpha = [\epsilonb]^\alpha \uk + \cdots \CR
Q \vc_{ij} = - [\gamma^\mu \tau_{\{i} \epsilon]_\alpha \partial_\mu \vcb^\alpha_{j\}} + \stfrac{1}{3} \delta_{ij} [\gamma^\mu \tau^k \epsilon]_\alpha \partial_\mu \vcb^\alpha_k + \cdots \hspace{7mm} Q \vcb_i^\alpha = [\epsilonb \tau^j ]^\alpha \vc_{ij} + [\epsilonb]^\alpha \vh_i + \cdots \CR
 Q \vk = -i[\gamma^\mu \epsilon]_\alpha \partial_\mu \vkb^\alpha+ \cdots \hspace{15mm} Q \vkb^\alpha = [\epsilonb]^\alpha \vh + \cdots \CR
Q \vh_i = -[\gamma^\mu \tau_i \epsilon]_\alpha \partial_\mu \vhb^\alpha + \cdots \hspace{10mm} Q \vh = -i[\gamma^\mu \epsilon]_\alpha \partial_\mu \vhb^\alpha+ \cdots \CR Q \vhb^\alpha = [\epsilonb]^\alpha \vh + [\epsilonb \tau^i ]^\alpha \vh_i + \cdots 
\end{gather}
where we have omitted the invariant tensor index $\xi$, and the index $\alpha$ stands for the $SU(2)$-Majorana representation. The transformations are linear in the physical sources, but non-physical terms can be products of the physical sources and the spurious ones that are coupled to BRST-exact operators. The transformations have the following general form
\bea Q [u^{\scriptscriptstyle \rm phys}] &=& [\epsilon] [u^{\scriptscriptstyle \rm phys}] +[\epsilon^2] [u^{\scriptscriptstyle \rm spur}]+ [\epsilon^2] [u^{\scriptscriptstyle \rm spur}][u^{\scriptscriptstyle \rm phys}] \CR
 Q [u^{\scriptscriptstyle \rm spur}] &=& 
[u^{\scriptscriptstyle \rm phys}] + [u^{\scriptscriptstyle \rm spur}] + [\epsilon] [u^{\scriptscriptstyle \rm spur}] + [\epsilon^2] [u^{{\scriptscriptstyle \rm spur}}] + [\epsilon^2] [u^{\scriptscriptstyle \rm spur}][u^{\scriptscriptstyle \rm phys}] + [\epsilon^2][u^{{\scriptscriptstyle \rm spur}\, 2}]\eea 
where the notation $[X^n]$ stands for covariant combinations of $X$ to the $n^{\rm th}$ power.

The transformations of the sources can be modified by radiative corrections through the generation of terms linear in the sources that are anomalies for the supersymmetry Slavnov--Taylor identity. The functionals of shadow number one and canonical dimension $4+ \frac{1}{2}$ that are invariant by the action of the linearized supersymmetry Slavnov--Taylor operator define infinitesimal deformations of the representation of supersymmetry given by the action of $Q$ on the sources, $Q \rightarrow Q +\varepsilon Q^{\scriptscriptstyle (1)}$ . Functionals, which can be written as the action of the linearized supersymmetry Slavnov--Taylor operator on another functional, define infinitesimal deformations that correspond to infinitesimal redefinitions of the sources. Thus the non-trivial elements of the cohomology correspond to possible deformations of the supermultiplets of composite operators that can arise in perturbation theory. In order that the composite operators be in representations of supersymmetry at the quantum level, the infinitesimal deformations of the supersymmetry representations must extend to finite representations. This property amounts to the existence of a deformed $Q^{\varepsilon}$ operator acting on the sources $u$ and $v$ as $Q^{\varepsilon} = Q + \varepsilon Q^{\scriptscriptstyle (1)} + \mathcal{O}(\varepsilon^2)$ that squares to a pure derivative ${Q^{\varepsilon}}^{\, 2} = - i (\epsilonb \gamma^\mu \epsilon) \partial_\mu$. The functionals of shadow number $2$ and canonical dimension $5$ that are non-trivial cohomology elements of $\Q_{|\Sigma}$ define obstructions to the extension of the infinitesimal deformations to finite deformations of the representations. However, computing the whole cohomology of shadow number $2$ is a very difficult problem. We will assume here that none of the corresponding anomalies arise in perturbation theory and that supersymmetry is preserved at the quantum level. Nevertheless, we will take care of the possible quantum deformations of the supermultiplets of composite operators at the quantum level associated the the cohomological class of shadow number $1$ in the complex of functionals linear in the sources. 

One can verify that the transformations of the sources $u$ and $v$ with an upper left index $C$ cannot be modified in a non-trivial way, because of their $U(1)_A \times SU(2)_R$ representations. Thus, the composite operator $ \trace [ \phi - \gamma_5 \phi ] \lambda$ does not mix with other operators by renormalization, and has the same anomalous dimension as the operators $\stfrac{1}{2}\trace \scal{ \phi_5^{\, 2} - \phi^2 } $ and $ \trace \phi \phi_5$, that is zero. In the same way, the operators $\stfrac{1}{2} \scal{ h^i \spro_\repre h^j - \stfrac{1}{3} \delta^{ij} h^k \spro_\repre h_k }$ and $[h^i - \stfrac{1}{3} \tau^i h ] \spro_\repre \psi$ can not mix with other operators, and have the same matrix of anomalous dimensions
\bea 
 \CS \Bigl[ \, \stfrac{1}{2} \scal{ h^i \spro_\repre h^j - \stfrac{1}{3} \delta^{ij} h^k \spro_\repre h_k } \ipro \Gamma \, \Bigr] &=& {\gamma_{\repre}}^\reprd \Bigl[ \,\stfrac{1}{2} \scal{ h^i \spro_\reprd h^j - \stfrac{1}{3} \delta^{ij} h^k \spro_\reprd h_k } \ipro \Gamma \, \Bigr] \CR
 \CS \Bigl[ \, [h^i - \stfrac{1}{3} \tau^i h ] \spro_\repre \psi \ipro \Gamma \, \Bigr] &=& {\gamma_{\repre}}^\reprd \Bigl[ \, [h^i - \stfrac{1}{3} \tau^i h ] \spro_\reprd \psi \ipro \Gamma \, \Bigr] \label{anoD}\eea
Instead of considering complete supermultiplets, we can also consider the insertion of composite operators through supersymmetry parameter dependent functions of the fields. It is particularly interesting to consider the coupling of non-trivial cocycles of the descent equations to external sources
\be \int \Scal{ u_0^0 \L_4^0 + u^{-1}_{1\, \wedge} \L_3^1 + u^{-2}_{2\, \wedge} \L_2^2 + u^{-3}_{3\, \wedge} \L_1^3 + u^{-4}_4 \L_0^4 } = \int \tilde{u}_{\, \wedge} \tilde{\L} \ee 
which we have written in the right-hand-side as the integral of the exterior product of the two extended forms $\tilde u$ and $\tilde \L$, with the Berezin prescription that only the top $4$-form gives rise to a non-zero integral. Integrating by part with respect to the extended differential $d + Q + i_{i(\epsilonb \gamma \epsilon)}$, it follows that the sources must transform as the components of a cocycle
\be (d + Q + i_{i(\epsilonb \gamma \epsilon)}) \, \tilde{u} = 0 \ee
in order that the action satisfies the supersymmetry Slavnov--Taylor identity. The same computation gives that the only invariant counter-terms, corresponding to mixing between operators by renormalization, are given by coupling cocycles to the sources. 

As a matter of fact, the representation of the differential $Q$ on the sources $u_p^{-p}$ is not modified by quantum corrections. This comes from the fact that the non-trivial cohomological elements of the linearized supersymmetry Slavnov--Taylor operator that are linear in the sources are in one to one correspondence with the non-trivial cocycles. These cocycles are themself in one to one correspondence with the functionals of the fields of canonical dimension $4+\frac{1}{2}$, linear in the supersymmetry parameter, which are non-trivial cohomological elements of the linearized supersymmetry Slavnov--Taylor operator. Such cohomological elements correspond to the consistent anomalies of the Slavnov--Taylor identities, which we know to do not exist.

The components with form-degree zero of the cocycles associated to the pure Yang--Mills action and the Chern character are linear combinations of the \textonehalf BPS operators $\stfrac{1}{2}\trace \scal{ \phi_5^{\, 2} - \phi^2 } $ and $ \trace \phi \phi_5$, and thus cannot mix between themself. Moreover, their components of form-degree one are linear combinations of the \textonehalf BPS operator $ \trace [ \phi - \gamma_5 \phi ] \lambda$ and thus these cocycles can only receive corrections from trivial cocycles. As a result, the pure Yang--Mills action and the Chern Character have no anomalous dimension, and can only admit BRST-exact invariant counter-terms
\be \CS \Bigl[ \, \int d^4 x \, \L_{\rm YM} \ipro \Gamma\, \Bigr] = \ins{1} \hspace{10mm} \CS \Bigl[ \, \int \trace F_{\, \wedge} F \, \ipro \Gamma \, \Bigr] = \ins{2} \ee 
This result is trivial for the Chern character at the perturbative level, since it is zero in this case. 

The components of form-degree zero of the hypermultiplet cocycles are just zero, and therefore, they can only mix between themself or with trivial cocycles. Moreover, the components of forms degree one, are the \textonehalf BPS operators $[h^i - \stfrac{1}{3} \tau^i h ] \spro_\repre \psi$, and thus
\be \CS \Bigl[ \, \int d^4 x \, \L_{\rm H}^\repre \ipro \Gamma\, \Bigr] = {\gamma^\repre}_\reprd \Bigl[ \, \int d^4 x \, \L_{\rm H}^\reprd \ipro \Gamma\, \Bigr] + \ins{3} \ee
with the same matrix of anomalous dimension as in equation (\ref{anoD}).

\section{Cancellation of the $\beta$ function}
Classically, we have the identity 
\be \frac{\partial \Sigma}{\partial g} = - \frac{2}{g^3} \int d^4 x \, \Scal{ \L_{\rm YM} + \L_{\rm H} } + \frac{1}{g} \S_{|\Sigma} \int d^4 x \, \scal{ \overline{G(\epsilon)} \spro \psiqs } \ee
Using the quantum action principle and the Ward identities, this identity implies the following equation for the generating functional of one-loop irreducible graphs \cite{Sorella}
\be \frac{\partial \Gamma}{\partial g} = - \frac{2 a(g) }{g^3} \Bigl[ \, \int d^4 x \L_{\rm YM} \ipro \Gamma \Bigr] - \frac{2 b_\repre (g) }{g^3} \Bigl[ \, \int d^4 x \L^\repre_{\rm H} \ipro \Gamma \Bigr] + \ins{4} \ee
Then, using the identity
\be \Bigl[ \CS , \frac{\partial \, }{\partial g} \Bigr] \mathscr{F} = - \frac{\partial \beta }{\partial g} \frac{\partial\mathscr{F} }{\partial g} + \S_{|\mathscr{F}} \mathscr{G} \ee
where $\mathscr{F}$
 is any given functional, one obtains the following equation \begin{multline}
 \frac{\partial \, }{\partial g} \biggl( \frac{ 2 a(g) \beta}{g^3} \biggr) \ \Bigl[ \, \int d^4 x \, \L_{\rm YM} \ipro \Gamma\, \Bigr] \\*+ \Biggl( \frac{\partial \, }{\partial g}\biggl( \frac{ 2 b_\repre (g) \beta}{g^3}\biggr) + \frac{ 2 b_\reprd (g) }{g^3} {\gamma^\reprd}_\repre \Biggl) \ \Bigl[ \, \int d^4 x \, \L_{\rm H}^\reprd \ipro \Gamma\, \Bigr] = \ins{5} \end{multline}

One can choose correlation functions in such way that all these terms are linearly independent. Thus, we get two differential equations
\be \frac{\partial \, }{\partial g} \biggl( \frac{ a(g) \beta}{g^3} \biggr) = 0 \hspace{10mm} \frac{\partial \, }{\partial g}\biggl( \frac{ b_\repre (g) \beta}{g^3}\biggr) + \frac{ b_\reprd (g) }{g^3} {\gamma^\reprd}_\repre = 0 \ee
The first equation implies that the $\beta$ function is proportional to its fist order $g^3$ component, and is thus zero to all orders of perturbation theory.

Inserting its solution in the second equation, we obtain that 
\be b_\reprd(g) \, {\gamma^\reprd}_\repre= 0 \ee
Therefore the \textonehalf BPS operators $\stfrac{1}{2} \scal{ h^i \spro h^j - \stfrac{1}{3} \delta^{ij} h^k \spro h_k } $ and $[h^i - \stfrac{1}{3} \tau^i h ] \spro \psi $ have anomalous dimensions equal to zero.
\be
 \CS \Bigl[ \, \stfrac{1}{2} \scal{ h^i \spro h^j - \stfrac{1}{3} \delta^{ij} h^k \spro h_k } \ipro \Gamma \, \Bigr] = 0 \hspace{10mm} 
 \CS \Bigl[ \, [h^i - \stfrac{1}{3} \tau^i h ] \spro \psi \ipro \Gamma \, \Bigr] = 0 \ee
 
 \section{Conclusion}
The validity of the superconformal Ward identities would imply that all the \textonehalf BPS primary operators of the hypermultiplet have vanishing anomalous dimensions \cite{Dobrev}. In this publication, we have demonstrated that supersymmetry Ward identities imply, independently of the regularization scheme, that all the \textonehalf BPS primary operators of the vector multiplet, the \textonehalf BPS primary operators $\stfrac{1}{2} \scal{ h^i \spro h^j - \stfrac{1}{3} \delta^{ij} h^k \spro h_k } $ and the first descendent $\trace [\phi - \gamma_5 \phi_5 ] \lambda$ and $[h^i - \stfrac{1}{3} \tau^i h ] \spro \psi $ have vanishing anomalous dimension to all order in perturbation theory. Moreover, the $\beta$ function is zero at all order of perturbation theory. This gives further support on the conjecture that $\N=2$ supersymmetry implies the superconformal invariance in a Yang--Mills theory with vanishing one-loop $\beta$ function. 
 
The main needed addition to our earlier proof of the cancellation of the $\beta$ function in the maximally supersymmetric theory has been the proof that, in the $\N=2$ theory, not only the last descendent of the pure Yang--Mills lagrangian is a protected operator, but also the last to the last. Moreover, our proof is expressed in terms of fields in fully covariant representations of the supersymmetric theory in Minkowski spacetime, as compared to the situation of the works \cite{Sorella,beta,BPS,descent}, where twisted variables were used in euclidean space.

\section*{Acknowledgments} 
We thanks very much E.~Rabinovici for having suggested us to extend our proof \cite{beta} to the non maximally supersymmetric cases. We are also grateful to K.~Stelle for very interesting discussions. 

This work was partially supported under the contract ANR(CNRS-USAR) \\ \texttt{no.05-BLAN-0079-01}.


\end{document}